\documentclass[aps,prd,preprintnumbers,showpacs]{revtex4}
\usepackage[dvips]{graphicx}
\begin{document}

%
%

\preprint{Nisho-4-2012}
\title{Anomalous Gluon Production and Condensation in Glasma}
\author{Aiichi Iwazaki}
\affiliation{International Economics and Politics, Nishogakusha University,\\ 
6-16 3-bantyo Tiyoda Tokyo 102-8336, Japan.}   
\date{Aug. 25, 2012}
\begin{abstract}
The collinear color electric and magnetic fields have been discussed to
be produced
immediately after high energy heavy ion collisions. 
We discuss anomalous gluon production under the background gauge fields.
The gluons are Nielsen-Olesen unstable modes.
The production rate of the modes by Schwinger mechanism has recently been found
to be anomalously larger than the rate of quarks
or other stable gluons. Analyzing classical evolutions
of the modes with initial conditions given by vacuum fluctuations,
we find that their production makes the color electric field decay very rapidly.
The life time of the field is approximately given by the inverse
of saturation momentum in the collisions. 
We also show that the mode with zero momentum form a Bose condensate and
its gluon number density grows up to be of the order of $1/\alpha_g$.
After the saturation of the gluon number density,
the condensate melts into quark gluon plasma 
owing to nonlinear interactions in QCD.
\end{abstract}
\hspace*{0.3cm}
\pacs{12.38.-t, 24.85.+p, 12.38.Mh, 25.75.-q, 12.20.-m, 03.75.Nt  \\
Schwinger mechanism, Chiral Anomaly, Color Glass Condensate}
\hspace*{1cm}

\maketitle

\section{introduction}

It has recently been paid attention to how
color electric ( $E$ ) and magnetic ( $B$ ) fields produced in high energy 
heavy ion collisions\cite{cgc} decay by producing quarks and gluons.
They form quark gluon plasma ( QGP ).
One of the most efficient driving forces for the decay is 
the Schwinger mechanism\cite{sch,tanji,iwazaki}. 
The massless quark production rate is similar to
the gluon production rate in the mechanism. 
The difference simply comes from the difference in the color charges,
spins and statistics.
However, when the color magnetic field ( $B$ ) is present, 
the gluon production rate becomes anomalously larger than
the quark production rate. 
This important fact has recently been shown by Itakura and Tanji\cite{ita}.
They have found that the gluon production rate
is proportional to $\exp(\pi B/E)$, while the quark production rate 
is proportional to $\exp(-\pi B/E)$ or unity.

As is well known, the gluons can have imaginary energies under the color magnetic field.
Namely, the states with magnetic moments parallel to $\vec{B}$ in
the lowest Landau level are unstable; their energies ( $\propto \sqrt{p_3^2-gB}$ ) are imaginary
when $p_3^2 < gB$,
where $p_3$ is a momentum component in the direction parallel to $\vec{B}=(0,0,B)$ 
( $g$ is a gauge coupling constant ).
Thus, their amplitudes exponentially grow with time.
The amplitudes of the other states of the gluons with real energies 
simply oscillate with time.
The presence of such unstable modes ( Nielsen-Olesen unstable modes\cite{nielsen} ) 
has prevented us to properly quantize
the gluons. Hence, the production of the Nielsen-Olesen unstable gluons by the Schwinger mechanism
could not be discussed. But the above authors pointed out that the production rate
can be obtained. This is because
the unstable modes become stable
owing to the acceleration by the electric field.
The momentum $p_3(t)=\int^t dt' \,gE$ becomes large sufficiently 
such that the imaginary energy becomes real for $p_3^2(t)>(gB)^2$. 
Consequently they have derived the gluon production rate.

The importance of their result is that the production rate, in other words,
the occupation number ( $\propto \exp(\pi B/E)$ ) of the gluon state 
is much larger than unity. 
This should be compared with the typical occupation number ( $\propto \exp(-\pi B/E)$ )
associated with other stable gluons or massless quarks\cite{tanji}. 
It is much less than unity.
The large gluon production rate leads to 
the rapid decay of the electric field. In particular, the decay process accelerates 
because the production rate $\propto \exp(\pi B/E)$ increases more
as the electric field $E$ becomes weaker. Hence, when we discuss the decay, 
it is important 
to include the back reaction
of the electric field to the gluon production. Furthermore, 
we may classically analyze the production of the unstable gluons
because of the large occupation number of the gluons.
In the classical treatment, we can easily take into account
the back reaction of the electric field to the gluon production. 
The back reaction has not yet been
discussed by the above authors\cite{ita}.

In this paper, 
by assuming that the unstable modes are initially produced by vacuum fluctuations, 
we analyze the classical evolution of the modes and the color electric field.
We find that the electric field rapidly decays owing to 
the acceleration of the large amount of the unstable gluons.
The life time of the electric field is approximately given by 
$Q_s^{-1}$ where $Q_s$ denotes saturation momentum in high energy heavy ion collisions.
( We assume in this paper that $gB$ and $gE$ is of the order of $Q_s^2$. )
We show that the color electric current carried by the gluons
is much larger than that of the quarks 
when they are produced by the Schwinger mechanism.
Thus,
the effect of the quarks on the decay of the electric field 
is negligible compared with that of the gluons.
We also show that the unstable gluons form a Bose-Einstein condensate, 
because of the large occupation number of the gluons. 
We find that the number density of the gluons in the condensate
grows up to be of the order of $1/g^2$.
After the formation of the condensate with such a large occupation number of gluons, 
it would melt into QGP with the equipartition of the momentum
by the nonlinear interactions of QCD.

In the next section \ref{2} we briefly review the Nielsen-Olesen unstable modes and discuss anomalous
production of the modes. In the section \ref{3},
we discuss field configurations of the Nielsen-Olesen unstable modes and
find basic equations governing the temporal behaviors of the modes 
as well as the color electric field. In the section \ref{4},
we find that the gluons of the modes are dominantly
produced and their production leads to the rapid decays of the electric field.
In the section \ref{5} we show that the Bose-Einstein condensation of the gluons
arises in which the gluon number density grows up to be of the order of $1/g^2$. We discuss
that the condensation with such large occupation number would melts 
owing to nonlinear interactions of the gauge fields.
In the final section \ref{6} we summarize our results.

\section{Nielsen-Olesen unstable modes}
\label{2}


We first explain our formalism and briefly review the Nielsen-Olesen unstable modes.
We also explain the anomalous production of the unstable modes 
under the electric field. 
We consider SU(2) gauge theory with
the background color electric and magnetic fields given by
$\vec{E}_a=\delta_{a,3}(0,0,E)$ and $\vec{B}_a=\delta_{a,3}(0,0,B)$.
They are supposed to be spatially homogeneous and
collinear both in the real and color spaces.
The gauge fields are represented by the gauge potential $A_{\mu}\equiv A_{\mu}^{a=3}$.
Under the background fields, the gauge potentials $\Phi_{\mu}\equiv (A_{\mu}^1+iA_{\mu}^2)/\sqrt{2}$ 
perpendicular to $A_{\mu}^3$ behave
as charged vector fields. When we represent SU(2) gauge potentials $A_{\mu}^a$ using 
the variables $A_{\mu}$ and $\Phi_{\mu}$, Lagrangian of SU(2) gauge potentials is
written in the following,

\begin{equation}
\label{L}
L=-\frac{1}{4}F_{\mu,\nu}^2-\frac{1}{2}|D_{\mu}\Phi_{\nu}-D_{\nu}\Phi_{\mu}|^2
-ig(\partial_{\mu}A_{\nu}-\partial_{\nu}A_{\mu})\Phi^{\dagger \mu}\Phi^{\nu}
+\frac{g^2}{4}(\Phi_{\mu}^{\dagger}\Phi_{\nu}-\Phi_{\nu}^{\dagger}\Phi_{\mu})^2,
\end{equation} 
with $F_{\mu,\nu}=\partial_{\mu}A_{\nu}-\partial_{\nu}A_{\mu}$ and 
$D_{\mu}=\partial_{\mu}-igA_{\mu}$, where we used a gauge $D_{\mu}\Phi^{\mu}=0$.
The gauge field $A_{\mu}$ represents both the background gauge field $A_{\mu,b}$ 
and fluctuations $\delta A_{\mu}$.
We find that the fields $\Phi_{\mu}$ represent charged vector fields
with the anomalous magnetic moment described by the term 
$-ig(\partial_{\mu}A_{\nu}-\partial_{\nu}A_{\mu})\Phi^{\dagger \mu}\Phi^{\nu}$. 
Therefore, it is easy to see that when the background magnetic field 
$B=\partial_1A_{2,b}-\partial_2A_{1,b}$ is present,
but $E=0$, 
the particles represented by the fields $\Phi$
occupy the Landau levels and interact with each other through the term
$\frac{g^2}{4}(\Phi_{\mu}^{\dagger}\Phi_{\nu}-\Phi_{\nu}^{\dagger}\Phi_{\mu})^2$.  
The energies of the states in the Landau levels denoted by integer $N\ge 0$ are 
given by $E_N=\sqrt{2gB(N+1/2)\pm 2gB+p_3^2}$, where $\pm$ denotes magnetic moment
parallel ( $-$ ) or anti-parallel ( $+$ ) to $\vec{B}$.

Among them we notice the states 
in the lowest Landau level ( $N=0$ ) with the magnetic moment parallel to $\vec{B}$. 
Their energies can be imaginary; $E_{N=0}=\sqrt{p_3^2-gB}$. 
Thus, the modes 
with the imaginary energies exponentially increase or decrease with time. That is,
the field $\Phi$ representing the modes evolves with time such that 
$\Phi\propto \exp(-iE_{N=0}t)=\exp(\pm |E_{N=0}|t)=\exp(\pm |\sqrt{gB-p_3^2}|\,t)$.
The states are called as Nielsen-Olesen
unstable modes. In particular, the mode with $p_3=0$ increases or decreases most rapidly.
The presence of such unstable modes implies the instability of the
vacuum state, i.e. $\langle\Phi\rangle=0$, when the background color magnetic field $B$ is present.
This is similar to the case that the state $\psi=0$ is unstable in a model of
a complex scalar field with the double well potential
$-m^2|\psi|^2+\frac{\lambda}{2} |\psi|^4$ ( $m^2>0$ ). 
In this model unstable modes exist around the state $\psi=0$ and exponentially grow
such as $\psi\propto \exp(t\sqrt{m^2-\vec{p}\,^2})$ 
where $\vec{p}$ denotes a momentum. 
Similarly the background gauge fields involving color magnetic fields are unstable.
Indeed, the classical simulations\cite{ven} have been performed
to show the instability of the states with the background gauge fields.
The instability in the simulation
has been discussed\cite{instability,instability2} to be
caused by the Nielsen-Olesen unstable modes: Their amplitudes exponentially
grow and then saturate when nonlinear interactions are effective
owing to the growth of the amplitudes.


The spontaneous production\cite{iwa} of the unstable modes ( or gluons ) is caused by the color
magnetic field. Thus, it is not the Schwinger mechanism.
When we analyze the Schwinger mechanism of the unstable modes, the presence of the
imaginary energy was an obstacle because we cannot properly quantize the modes.
But, we should note that in order to obtain the gluon production rate 
by the Schwinger mechanism\cite{tanji}, 
we only need in-state ( in infinite past ) and out-state
( in infinite future ) under the electric field. Although the Nielsen-Olesen modes are unstable,
they are stable in the infinite past and future in the presence of the
electric field. This is because the square of the momentum $p_3(t)=\int^t dt' \,gE$ is sufficiently large
in the past and future such as $p_3^2(t)-gB>0$.
The stability of the unstable modes in the infinite past and future  
allows us to estimate the production rate\cite{ita} of the modes.

In general, charged fields oscillate with the frequency $\propto gEt$ in the past
( $t\to -\infty$ ) and future ( $t\to +\infty$ ) under the electric field $E$.
The frequency is real and depends on time. They oscillate even in the period
between the past and the future.
In this case the production rate of the fields is less than
unity; it is 
proportional to, for example, $\exp(-\pi B/E)$ when the magnetic field is present.
But, in the case of the unstable modes 
they pass a period in which their frequency becomes imaginary 
so that their amplitudes exponentially grow. Before and after the period
they simply oscillate with the real frequency. 
In other words, the modes oscillate and their amplitudes smoothly change in the far past,
but once they enter the period, their amplitudes exponentially grow with time.
After passing the period, they
oscillate again but with much larger amplitudes than those
before passing the period.
These behaviors are peculiar to the unstable modes. In particular,
the exponential growth\cite{ita} of the amplitudes leads to 
the anomalous production rate $\propto \exp(\pi B/E)$.

We may understand a naive physical reason why the rate increases more as the electric 
field becomes weaker.  The large production rate
comes from the fact that the modes pass the period in which they exponentially grow.
They stays in the period approximately for $\Delta t =\sqrt{gB}/gE$ because
the momentum increases in the period such that $\Delta p_3 =\Delta t\, gE=\sqrt{gB}$.
Hence, the amplitude grows by $\exp(\sqrt{gB}\Delta t)=\exp(B/E)$.
Weaker electric field causes the longer stay in the period 
and hence the larger growth of the amplitude.
Although it is a rough estimation, it explains why the gluon production rate
$\propto \exp(\pi B/E)$ becomes large as the electric field becomes weak.


\vspace{0.2cm}
Hereafter we take only the unstable modes and analyze their production under the electric field.
The unstable modes are described by the field $\Phi\equiv (\Phi_1+i\Phi_2)/\sqrt{2}$ and
are governed  by the following Hamiltonian

\begin{equation}
H=\int d^3x \Big(\frac{1}{2}(\partial_0A_{3,b})^2+|\partial_0\Phi|^2+
|(\vec{\partial}-ig\vec{A_b})\Phi|^2-2gB|\Phi|^2\Big )=\int d^3x 
\Big(\frac{1}{2}(\partial_0A_{3,b})^2+|\partial_0\Phi|^2+
|(\partial_3-igA_{3,b})\Phi|^2-gB|\Phi|^2\Big ),
\end{equation}
where we neglected the nonlinear interactions 
$\frac{g^2}{4}(\Phi_{\mu}^{\dagger}\Phi_{\nu}-\Phi_{\nu}^{\dagger}\Phi_{\mu})^2$ in eq(\ref{L}). 
The color electric and magnetic fields are given such that 
$\vec{E}=\partial_0 \vec{A_b}=(0,0,\partial_0A_{3,b})$
and $\vec{B}=\vec{\partial}\times \vec{A_b}=(0,0,B)$.
The nonlinear interactions are not effective
as long as the amplitude $\Phi$ is small. When the field $\Phi$ grows large
such as the nonlinear interactions are effective,
the unstable modes couple with the other modes in higher Landau levels
as well as themselves.

We note that the last term in the Hamiltonian represents a negative potential.
When the magnetic field forms a flux tube, the term represents a negative potential
with finite width given by the width of the flux tube. Whether or not
the field $\Phi$ possesses unstable modes depends on the existence of
the states trapped in the negative potential.

\vspace{0.2cm}
It is interesting to see the analogy between our model and the model of the complex scalar field
with the double well potential mentioned above. The model describes Cooper pairs
condensates. Thus, the decay of the color electric field corresponds to the decay
of an external electric field imposed on superconductors. 
The question is how fast the external electric field decays, 
just after a metal in a normal state
is supercooled below the critical temperature
at which 
the normal state ( $\langle\psi\rangle=0$ )
and superconducting state ( $\langle\psi\rangle=\sqrt{m^2/\lambda}$ ) are separated.  
The normal state $\psi=0$
decays by producing the Cooper pairs $\psi$, which condense to form
the state $\langle\psi\rangle=\sqrt{m^2/\lambda}$. Since they are accelerated by
the electric field, the electric field loses its energy and vanishes.

\section{production of Nielsen-Olesen unstable modes}
\label{3}

Because the production rate is much larger than unity ( this implies that
the occupation number in a state is much larger than unity ),
the production of the unstable gluons may be classically analyzed. 
Then, we can easily take into account the back reaction of the electric field
to the gluon production. In this section we will formulate basic equations
governing the back reaction.

First, we discuss the assumption
that the background gauge fields are homogeneous in the transverse plane perpendicular
to the collinear fields $\vec{B}$ and $\vec{E}$. 
When the unstable modes are excited, they destroy the homogeneity
because of the localization of the wave functions of the modes.
 
\begin{equation}
\phi\equiv (x_1-ix_2)^n\exp(-\frac{gB|z|^2}{4}+ip_3x_3),
\end{equation}
with $z\equiv x_1+ix_2$ and integer $n\ge 0$
where we used a gauge potential $\vec{A_b}=(-Bx_2/2,Bx_1/2,0)$.
The effect of the back reaction induces the inhomogeneity in the background gauge fields;
the currents carried by the modes are not homogeneous so that
the background gauge fields affected by the currents are also not homogeneous.

But, by taking the appropriate linear combination of the unstable modes we
can form almost homogeneous field configurations
in the transverse plane.
Then, their currents are also almost homogeneous.
Such field configurations are given by,

\begin{equation}
\label{sum}
\Phi=\sum_{l=1\sim N}\phi_l(\vec{x}), \quad \phi_l(\vec{x})
=\int dp_3 \,c(p_3)\exp(-\frac{gB|z-z_l|^2}{4}+ip_3x_3),
\end{equation} 
with $c(p_3)$ is a dimensionless function of the longitudinal momentum $p_3$ and $z_l=x_{1.l}+ix_{2,l}$,
where each component $\phi_l$ satisfies the condition, 
$\phi_l\phi_{l'}\simeq \delta_{l,l'}\phi_l^2$ 
because we impose that $|z_l-z_{l'}|\geq l_B\equiv \frac{1}{\sqrt{gB}}$.
Namely, a configuration $\phi_l$ is separated with the nearest neighbors 
approximately by the distance $l_B$.
Furthermore, we assume that the area $L^2$ 
of the transverse plane is given by $L^2=Nl^2_B$.
Thus, we find that the field configuration $\Phi$ is approximately uniform in the transverse space.
This kind of the configuration of the unstable modes was analyzed\cite{ninomiya} to discuss so called
"spaghetti vacuum". 

Using the field configuration, we rewrite the Hamiltonian of the unstable modes,

\begin{equation}
H=\int d^3x \Bigg(\frac{1}{2}(\partial_0A_b)^2+|\partial_0\Phi|^2+|(i\partial_3-gA_b)\Phi|^2
-gB|\Phi|^2\Bigg)\simeq N\int d^3x 
\Bigg(\frac{1}{2}(\partial_0A_b)^2+|\partial_0\phi|^2+|(i\partial_3-gA_b)\phi|^2
-gB|\phi|^2\Bigg),
\end{equation}
with $\phi=\int dp_3 \,c(p_3)\exp(ip_3x_3-gB|z|^2/4)$,
where the color electric field is given by $E=\partial_0A_b$ with 
the homogeneous gauge potential $A_b\equiv A_{3,b}$. 
The Hamiltonian describes the unstable modes under the homogeneous background electric and magnetic
fields.
The first term represents the energy of the electric field
and the other terms represent the energy of the unstable modes.
We can see that the last term with the magnetic field $gB$ 
represents a negative potential for the unstable modes $\phi$.
Thus, it gives rise to the imaginary energy of the field 
$\phi\propto \exp(-i\epsilon t)$ with $\epsilon^2=-gB<0$. 

If the magnetic field forms a flux tube with a finite width,
it gives a negative potential with the finite width. Thus, if the field is trapped by 
the potential, the energy $\epsilon $ can be imaginary, 
but its absolute value
is smaller than $\sqrt{gB}$; it depends on the width of the tube.
When the width is infinite, the energy $\epsilon $ is given by $\sqrt{-gB}$.
On the other hand, when the width is finite,  
the absolute value of energy $\epsilon $
becomes smaller, as the width becomes smaller.
The flux tubes of the background gauge fields are more realistic field configurations
produced in high energy heavy ion collisions than the homogeneous ones under consideration.
Because the gauge fields are homogeneous in the longitudinal direction,
they can be viewed as an ensemble of electric and magnetic flux tubes with various widths.
Based on the view, we have discussed\cite{instability,instability2} 
the results of the numerical simulations\cite{ven}. 
Although the flux tubes are realistic ones,
it is meaningful to analyze the anomalous gluon production in the homogeneous 
gauge fields, in order to see physical essences of the production.

\vspace{0.2cm}
We proceed to analyze the back reaction.
For the purpose, we decompose the field $\phi(\vec{x})$ into the 
components of the momentum eigenstate,

\begin{equation}
\phi=\frac{1}{\sqrt{L^3}}\sum_{\vec{p}}\phi_p \exp(i\vec{p}\,\vec{x})
\end{equation} 
with
\begin{equation}
\label{moment}
\phi_p=\frac{8\pi^2c(p_3)}{gBL^{3/2}}\exp(-\frac{p_T^2}{gB})
\end{equation}
where the transverse momentum $p_T$ is given such that $p_T^2\equiv p_1^2+p_2^2$.
Then, it follows that

\begin{equation}
H=L^2\Bigg(\frac{L}{2}(\partial_0 A_b)^2+\frac{1}{l_B^2}\sum_{\vec{p}}(\,|\partial_0 \phi_p|^2+
|(p_3+gA_b)\phi_p|^2-gB|\phi_p|^2\,)\Bigg),
\end{equation}
where we have used the following formulas,

\begin{equation}
\int d^3x\exp(i\vec{p}\vec{x})=(2\pi)^3\delta^3(p)=L^3\delta^3_{p,0}.
\end{equation} 

The equations of motions of the fields $\phi_p$ and the gauge field $A_b$ are respectively given by 

\begin{equation}
\label{eqm}
\partial_0^2\phi_p=gB\phi_p-(p_3+gA_b)^2\phi_p \quad \mbox{and} \quad
L\partial_0^2A_b=-\frac{2g}{l_B^2}\sum_{\vec{p}}(p_3+gA_b)|\phi_p|^2 
\end{equation}  
where the second equation represents 
a Maxwell equation $\partial_0E=-J$ with the current $J=\frac{2g}{l_B^2}\sum_{\vec{p}}(p_3+gA_b)|\phi_p|^2$. 
It describes how the electric field changes by the effect of the current $J$.

We should point out that 
the largest amplitude of the unstable modes is given by the mode with 
the vanishing transverse
and longitudinal momentum, that is, the mode with $p_T=0$ in eq(\ref{moment})
and $p_3+gA_b=0$ in eq(\ref{eqm}). The mode grows most rapidly
compared with the other modes with $p_3+gA_b\neq 0$. Here, the momentum 
$p_3(t)\equiv p_3+gA_b(t)=p_3+g\int_0^t dt'E(t')$ denotes that of the mode 
with the initial momentum $p_3(t=0)=p_3$.
We naively expect from the similarity to the model of the complex scalar field
that the mode with zero momentum form a stable Bose condensate. But
as we discuss later, 
although the mode form a Bose condensate,
the condensate becomes unstable when its amplitude grows up to of the order of $1/g$. 

\vspace{0.2cm}
In order to solve the equations we need to impose initial conditions.
The initial condition of the electric field is given such that $E(t=0)=\partial_0A_b(t=0)=E_0$
and $A_b(t=0)=0$. This corresponds to the fact that we switch on the electric field $E=E_0$ at $t=0$.
In other words, we consider the situation that high energy heavy ion collisions occur at $t=0$ 
and the color backgroung gauge fields are produced at the instance. 

How should we choose initial conditions of the field $\phi$ ?
Before the collisions, there are no color electric and magnetic fields.
The gluons with small $x$ form color glass condensates in nuclei. Just after the collisions
the gluons form the coherent gauge fields $E$ and $B$, but
there is no classical field $\phi$. Thus, we may naively choose the initial conditions
such that $\partial_0\phi(t=0)=0$ and $\phi(t=0)=0$. But these initial conditions
lead to the trivial result; $\phi(t)=0$ for any time $t>0$.
Therefore, we need to find other appropriate initial conditions. 
As we explained in the previous section, the unstable modes are spontaneously generated 
owing to the instability of the state with the homogeneous magnetic field.
Thus, it is reasonable to take the initial conditions given by the vacuum fluctuations,

\begin{equation}
\label{ini}
\phi_p(t=0)\equiv\sqrt{\langle \hat{\phi}_p^2 \rangle} 
\quad \mbox{and} \quad \partial_t\phi_p(t=0)\equiv \sqrt{\langle (\partial_t\hat{\phi_p})^2\rangle}
\end{equation}
where $\hat{\phi}_p$ denotes the momentum component of 
the free massless scalar field $\hat{\phi}$ with no
background fields. The state $|\,\,\rangle $ 
represents the vacuum state $\langle \hat{\phi}\rangle=0 $. 
The vacuum can be represented by the following wave functionals

\begin{equation}
\label{vac}
W({\phi_p})\propto \exp\Big(-\sum_{\vec{p}}|p||\phi_p|^2\Big) \quad \mbox{and} \quad
W({\partial_0\phi_p})\propto \exp\Big(-\sum_{\vec{p}}\frac{|\partial_0\phi_p|^2}{|p|}\Big),
\end{equation} 
with $|p|\equiv \sqrt{p_T^2+p_3^2}$.
Namely, we assume that the initial conditions are given by
the vacuum fluctuations in the vacuum without $E$ and $B$. 

It apparently seems that we should use vacuum fluctuations
when the color magnetic field
is present.
But, the vacuum fluctuations cannot be defined when $B$ is present,
because there is no stable vacuum owing to the Nielsen-Olesen instability. 
This is the reason why we use the vacuum
fluctuations in the vacuum without $E$ and $B$.  
Furthermore, what we need to obtain as an initial condition
is the longitudinal momentum distribution at $t=0$.
The transverse momentum distribution of the unstable modes has been
found since we are only concerned with the states in the Lowest Landau level.   
In this way we take into account the inital conditions given by vacuum fluctuations
in the vacuum without $E$ and $B$.
With regard to the initial conditions for the unstable modes, 
an intriguing research\cite{ven2} has been performed in expanding glasma. 
In the final section we make a comment on the relation between the initial condition
in our paper and the one dictated by the reference.

If we consider the case that the double well potential, 
$-m^2|\psi|^2+\frac{\lambda}{2} |\psi|^4$ ( $m^2>0$ ),
is added to massless scalar field at $t=0$, it is reasonable to think that the vacuum
fluctuations in the vacuum of the massless scalar field evolve to make a stable state 
$\langle\psi \rangle\neq 0$ just after the addition of the potential. 
The situation is similar to the case of the glasma.

\vspace{0.2cm}

Using the wave functionals, we shall find the distribution of the
Nielsen-Olesen unstable modes in the vacuum.
In order to do so,
we rewrite the field $\phi_p$ such that

\begin{equation}
\label{flu}
\phi_p=\frac{(\tilde{\phi}_1(p_3)+i\tilde{\phi}_2(p_3))\sqrt{\pi}}{(gB)^{3/4}L\,
\sqrt{T(p_3l_B)}}\exp(-\frac{p_T^2}{gB}) 
\end{equation} 
with dimensionless real functions $\tilde{\phi}_i(p_3)$. Then, the wave functionals are given by

\begin{equation}
\label{functional}
W({\tilde{\phi}})\propto 
\exp\Big(-\sum_{p_3}\frac{\big(\tilde{\phi}_1(p_3)\big)^2+\big(\tilde{\phi}_2(p_3)\big)^2}{4}\Big) \quad 
\mbox{and} \quad
W({\partial_0\tilde{\phi}})\propto 
\exp\Big(-\sum_{p_3}\frac{\big((\partial_0\tilde{\phi}_1)^2+(\partial_0\tilde{\phi}_2)^2\big)U(p_3l_B)}
{4gBT(p_3l_B)}\Big),
\end{equation}
where 
\begin{equation}
T(x)\equiv\int_0^{\infty} dy \sqrt{y+x^2}\,\exp(-2y) \quad \mbox{and} \quad
U(x)\equiv \int_0^{\infty} dy \frac{\exp(-2y)}{\sqrt{y+x^2}}.
\end{equation}

This is the distribution of the Nielsen-Olesen unstable modes in the vacuum
of the massless scalar field. That is the distribution of $\tilde{\phi}$.
We note that the distribution of the real part $\tilde{\phi}_1$ is identical to 
that of	the imaginary part $\tilde{\phi}_2(p_3)$. 
Using the distribution, we obtain 
the expectation values of the unstable modes,

\begin{equation}
\langle \tilde{\phi}_i(p_3)^2 \rangle=
\frac{\int d\tilde{\phi}_i|W(\tilde{\phi})|^2\tilde{\phi}_i^2}
{\int d\tilde{\phi}_i|W(\tilde{\phi})|^2}=1
\quad \mbox{and} \quad
\langle (\partial_0\tilde{\phi}_i(p_3))^2 \rangle=
\frac{\int d(\partial_0\tilde{\phi}_i)|W(\partial_0\tilde{\phi})|^2(\partial_0\tilde{\phi}_i)^2}
{\int d(\partial_0\tilde{\phi}_i)|W(\partial_0\tilde{\phi}_i)|^2}=\frac{gBT(p_3l_B)}{U(p_3l_B)},
\end{equation}
with $i=1,2$.

Therefore, we find that the equations of motion are given by 

\begin{equation}
\partial_0^2\tilde{\phi}(p_3)=gB\,\tilde{\phi}(p_3)-(p_3+gA_b)^2\tilde{\phi}(p_3) \quad \mbox{and}\quad
\partial_0^2 gA_b=-\frac{g^2\sqrt{gB}}{4\pi}\int_{-\infty}^{+\infty} 
dp_3 \frac{(p_3+gA_b)\tilde{\phi}(p_3)^2}{T(p_3l_B)}
\end{equation}
with $\tilde{\phi}\equiv \tilde{\phi}_i$ and the initial conditions  

\begin{equation}
A_b(t=0)=0, \quad \partial_0 A_b(t=0)=E_0 \quad \mbox{and} \quad
\tilde{\phi}(p_3,t=0)=1, \quad  \partial_0\tilde{\phi}(p_3,t=0)
=\sqrt{\frac{gBT(p_3l_B)}{U(p_3l_B)}},
\end{equation}   
where we used the relation $\sum_{p_3}=\frac{L}{2\pi}\int_{-\infty}^{+\infty} dp_3$.

We should stress that the initial conditions of $\tilde{\phi}$ was determined 
with the distribution of the unstable modes in the vacuum without $E$ and $B$.
The initial condition gives
the longitudinal momentun distribution $\tilde{\phi}$ at $t=0$.  
Namely, we have simply determined 
the dependence on $p_3$, i.e. $\tilde{\phi}(p_3)$ 
of the Nielsen-Olesen unstable modes $\phi_p$ in eq(\ref{flu})
by using the vacuum functional in eq(\ref{functional}) and used it as the initial condition.

Obviously both the equations of motion and initial conditions are independent on the system size $L$.
Furthermore, the color electric current $J$ is given by

\begin{equation}
J=\frac{g\sqrt{gB}}{4\pi}\int_{-\infty}^{+\infty} dp_3 \frac{(p_3+gA_b)\tilde{\phi}(p_3)^2}{T(p_3l_B)},
\end{equation}
which is also independent on $L$.
The current vanishes at $t=0$ because $T(x)=T(-x)$ and $A_b(t=0)=0$.
Using these equations of motion,
we can discuss the temporal behaviors of the electric field $E(t)=\partial_0A_b$ and the unstable modes
$\tilde{\phi}(p_3,t)$ by taking into account the back reaction of the electric field 
to the production of the modes.  

For the convenience, we write down the field $\phi$ of the unstable modes 
in terms of the variable $\tilde{\phi}$,

\begin{equation}
\phi=\frac{1}{\sqrt{L^3}}\sum_{\vec{p}}\phi_p \exp(i\vec{p}\,\vec{x})=
\frac{1}{\sqrt{L^3}}\sum_{\vec{p}}\,\frac{(\tilde{\phi}(p_3)+i\tilde{\phi}(p_3))\sqrt{\pi}}
{(gB)^{3/4}L\sqrt{T(p_3l_B)}}
\exp(-\frac{p_T^2}{gB})\exp(i\vec{p}\,\vec{x}),
\end{equation}
where $\tilde{\phi}(p_3)$ is a dimensionless real function.
We note that the vacuum fluctuation $\phi(t=0)$ of 
the unstable modes is of the order unity, while the
background gauge fields $B$ and $E(t=0)=E_0$ are of the order of $\sim O(1/g)$.


\section{numerical results}
\label{4}

Now we wish to discuss the production of the Nielsen-Olesen gluons and quarks. Especially,
we would like to discuss the ratio between the amount of the gluons and that of the quarks
produced by the electric field. In order to
discuss the amounts of the particles we compare
the color electric current of the quarks with that of the gluons.  
We show that the amount of the gluons is about a hundred times larger than that of the quarks.
As a result the color electric field rapidly decays owing to this anomalous gluon production.

First we derive a relevant equation describing the quark production
by the Schwinger mechanism. The equation has been previously derived\cite{iwazaki}.
We would like to explain it briefly.
We assume that the quarks are massless and they form a SU(2) doublet.
Then,
the color charges of the quarks coupled with $A_{\mu}$ are given by $g/2$ and $-g/2$.
Both of them possess their anti-particles with their charges
given by $-g/2$ and $g/2$, respectively. Therefore, we have four massless fermions;
a pair of the quarks ( $q_{+}$ and $q_{-}$ ) in a SU(2) doublet 
and their anti-quarks ( $\bar{q}_{+}$ and $\bar{q}_{-}$ ). 
The quarks ( $q_{+}$ and $\bar{q}_{-}$ ) have
the positive charge $g/2$ and the quarks ( $q_{-}$ and $\bar{q}_{+}$ ) 
have the negative charge $-g/2$. 
Their number densities are identical to each other because
a pair of a positive and a negative charged quarks is created at the same moment
under the electric field. 

It has been recently shown\cite{iwazaki,suganuma} that
the evolution of the number density $n_q$ of the massless fermions  
is governed by the chiral anomaly 
when collinear strong electric and magnetic fields are present.
In particular the anomaly equation becomes very simple
when the magnetic field is sufficiently strong
such that the particles produced occupy only 
the states with the lowest energy. 
Namely, the equation of the chiral anomaly is given by 

\begin{equation}
\label{chi}
\partial_0J_0^5=4\partial_0n_q=2\frac{(g/2)^2E(t)B}{2\pi^2},
\end{equation}
where we assumed the homogeneity of the chiral current 
in the transverse and longitudinal direction; $\vec{\partial}\vec{J}\,^5=0$.
The equality $\partial_0J_0^5=4\partial_0n_q$ in eq(\ref{chi}) comes from the fact that
all of the four fermions have the positive chirality when
$\vec{E}$ parallel to $\vec{B}$.
This is because the positive ( negative ) charged fermions are accelerated 
to the direction parallel ( antiparallel ) to $\vec{E}$ and their spins  
are pointed to the direction ( antiparallel ) parallel to $\vec{B}$ 
when they occupy the states in the lowest Landau level.

Obviously, the chiral anomaly in eq(\ref{chi}) describes how the number density $n_q$ evolves with time
under the effect of the electric and magnetic fields.
The electric field loses its energy owing to the acceleration of the quarks
as well as the gluons. Hence, we add the contribution of the quarks to the
Maxwell equation. Consequently, the equations describing the evolution of the numbers of the
quarks and the gluons as well as the evolution of the electric field are given by

\begin{eqnarray}
\label{eq}
\partial_0n_q&=&\frac{g^2E(t)B}{16\pi^2},\quad 
\partial_0^2\tilde{\phi}(p_3)=gB\,\tilde{\phi}(p_3)-(p_3+gA_b)^2\tilde{\phi}(p_3) \nonumber \\
\partial_0^2 gA_b &=&-4g^2n_q-\frac{g^2\sqrt{gB}}{4\pi}\int_{-\infty}^{+\infty} 
dp_3 \frac{(p_3+gA_b)\tilde{\phi}(p_3)^2}{T(p_3l_B)},
\end{eqnarray} 
where $4gn_q$ denotes the current of the four kinds of the massless quarks.
The initial conditions are given by

\begin{equation}
n_q(t=0)=0, \quad \tilde{\phi}(p_3,t=0)=1, \quad 
\partial_0\tilde{\phi}(p_3,t=0)=\sqrt{\frac{gB\,T(p_3l_B)}{U(p_3l_B)}},
\quad A_b(t=0)=0 \quad \mbox{and} \quad \partial_0A_b(t=0)=E_0 .
\end{equation}

By solving these equations we can see how the electric field vanishes 
owing to the production of the unstable modes and the quarks.
Furthermore, we can obtain the temporal behaviors of 
the electric current densities of the quarks $J_q=4gn_q$ 
and the gluons $J_g=\frac{g\sqrt{gB}}{4\pi}\int_{-\infty}^{+\infty} 
dp_3 \frac{(p_3+gA_b)\tilde{\phi}(p_3)^2}{T(p_3l_B)}$.  
Both of them vanish at $t=0$. After the electric field is switched on at $t=0$, the pair production
of the quarks arise and their electric current flows. Similarly, 
the gluons as the unstable modes are produced
as the vacuum fluctuations and their electric current flows along the electric field. 
Owing to the production of the quarks and the gluons, the electric field
decreases and vanishes at $t=t_c>0$. 
Hence we compare the electric current of the quarks with that of the gluons at 
$t=t_c$ when the electric field vanishes,

\begin{equation}
R(t=t_c)=\frac{J_g}{J_q}=\frac{\frac{g\sqrt{gB}}{4\pi}\int_{-\infty}^{+\infty} dp_3 
\frac{\big(p_3+gA_b(t=t_c)\big)\tilde{\phi}(p_3,t=t_c)^2}{T(p_3l_B)}}{4gn_q(t=t_c)},
\end{equation}
where all of the integration range of the momentum $p_3$ has been taken.
But the relevant modes we should take into account are the unstable modes.
The integration range should be limited to the range 
in which each mode $\tilde{\phi}(p_3,t)$ can exponentially increase.
For example, the modes $\tilde{\phi}(p_3,t)$ with $p_3>\sqrt{gB}$
never exponentially increases so that we should not include the modes.
On the other hand, the mode $\tilde{\phi}(p_3,t)$ 
with $|p_3+gE_0 t_c|>\sqrt{gB}$ and $p_3<0$ can pass the period
in which it exponentially increases until the electric field vanishes.
Hence, the rellevant integration range is 
approximately given such that
$|p_3+gE_0 t_c| >\sqrt{gB}$ for $p_3<0$ and $p_3<\sqrt{gB}$ for $p_3>0$.
As our analysis of the gluons is classical,
our estimation of the ratio is not rigorous. 
Thus, rigorously speaking, we do not know 
the appropriate integration range:
If we take all of the integration range, 
irrelevant modes to Nielsen-Olesen instability are taken into account.
Therefore, we may limit the integration range such as $|p_3|<\sqrt{gB}$, in order to see roughly 
how large amount of the gluons are produced compared with that of the quarks. 

In Fig.1 we show the temporal behaviors of the electric field and the ratio $R$ with the parameters
$g=B=E_0=1$. We have taken the integration range $|p_3|<\sqrt{gB}$ and
checked that the result does not change even if the integration range
$|p_3|<1.2\sqrt{gB}$ is taken.
We find that when the electric field vanishes, the electric current of the gluons 
is approximately
eighty times larger
than that of the quarks. Owing to this fact, the life time of the electric field
is much shorter than the life time only when  
the quark production is taken into account.
Actually, the above equations can be explicitly solved if the contribution of the gluons is
neglected, i.e. $\tilde{\phi}=0$. The solution of the electric field is given by

\begin{figure}[htb]
  \centering
  \includegraphics*[width=65mm]{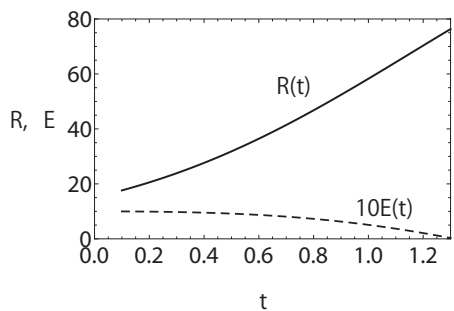}
  \caption{temporal behaviors of 
 the ratio $R(t)$ (solid line) and ten times electric field $10\times E(t)$ (dashed line)}
     \label{f1}
\end{figure}

\begin{equation}
E=E_0\cos\Bigl(\frac{\sqrt{g^3B}\,t}{2\pi}\Bigr).
\end{equation}
The solution represents a plasma oscillation\cite{tanji,iwazaki}.
Hence, the life time $t_c$ at which $E$ vanishes is given by $\pi^2/\sqrt{g^3B}$, which
is equal to $\pi^2\sim 10$ with $g=B=1$. It is roughly 8 times longer than the life time
shown in the figure. The life times are given by $1.3 Q_s^{-1}$ and $10 Q_s^{-1}$ respectively 
in the physical unit $Q_s^{-1}$; $Q_s$ ( $ =1\rm{GeV}\sim 2\rm{GeV} $ ) denotes saturation momentum
of high energy heavy ion collisions in RHIC or LHC.
In this way the decay of the electric field is mainly caused by the anomalous gluon production,
that is, the production of the unstable modes. The contribution of the quarks is negligible.
As we show in next section, the life time 
$t_c$ is of the order of $Q_s^{-1}$, while it is of the order of $Q_s^{-1}g^{-1}$ 
in the case of no gluon production. Thus, $R$ becomes larger as $g$ becomes smaller,
because the quark production is suppessed as $g$ becomes small.

\vspace{0.2cm}
The unstable modes are generated at $t=0$ by the vacuum fluctuations and are amplified by the magnetic field.
At the same time, they are accelerated by the electric field.
In Fig.2 we show the temporal behavior of the momentum distribution $\phi_p$ or 
$\tilde{\phi}(p_3,t)/\sqrt{T(p_3l_B)}$, that is, the growth of the amplitude with time. 
We can see that the vacuum fluctuations ( shown by the curve at $t=0$ in Fig.2 ) 
give the momentum distribution
symmetric in $p_3$ with the peak at $p_3=0$. 
The peak moves to points with negative momentum with time. 
This is because the gluons are accelerated by the electric field so that
the momentum with which the amplitude $\tilde{\phi}(p_3,t)/\sqrt{T(p_3l_B)}$ has
the largest growth rate is given by $p_3=-\int_0^t dt'gE(t')=-gA_b(t)<0$.

\begin{figure}[htb]
  \centering
  \includegraphics*[width=65mm]{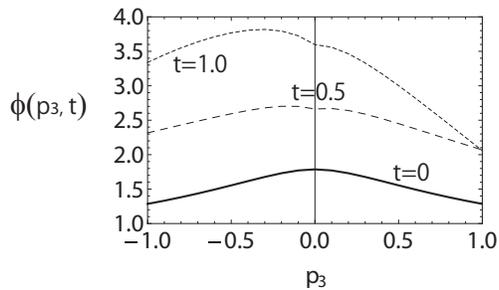}
  \caption{$\phi(p_3,t)\equiv\tilde{\phi}(p_3,t)/\sqrt{T(p_3l_B)}$ at
$t=0$ ( solid line ), $t=0.5$ ( dashed line ) and $t=1.0$ ( short dashed line) }
     \label{f2}
\end{figure}

We can see, however, from the figure that the peak is not located at $p_3=-\int_0^t dt'gE(t')$, but
at momentum near $p_3=0$ as long as $t$ is small. This is because the initial condition 
$\frac{\partial_0\tilde{\phi}(p_3,t=0)}{\sqrt{T(p_3l_B)}}=\sqrt{\frac{gB}{U(p_3l_B)}}$
becomes larger, as $|p_3|$ becomes larger. Thus, as long as $t$ is small, 
the peak stays at $p_3=0$ by the effect of the initial condition. 
The effect diminishes
with time. Indeed, the maximum of the amplitude approaches the value 
$\phi_p\big(p_3=-\int_0^t dt'gE(t')\big)$ as $t$ goes beyond $0.5$, while
it is given by $\phi_p(p_3\simeq 0)$ at $t<0.5$.

\section{gluon condensation}
\label{5}
We have mentioned the similarity between the Nielsen-Olesen unstable modes and unstable modes
in the model of the complex scalar field. The model describes Cooper pairs in superconductors.
Obviously, the unstable modes with zero momentum in the model 
form a Bose condensate $\langle\psi\rangle=\sqrt{m^2/\lambda}$ of the Cooper pairs.
That is, once the modes are produced, 
the amplitudes of the modes exponentially grow and take the value of the stable
state. They are stabilized by the nonlinear interaction $\lambda |\psi|^4$. 
Hence, it is natural to expect from the similarity 
that the unstable modes of the gluons may form a Bose condensate\cite{condense}. 

The unstable modes typically carry the momentum
$\vec{p}=(p_T\simeq \sqrt{gB},p_3\simeq \sqrt{gB})$. 
This is because the modes have the spatially transverse
extension given by $1/\sqrt{gB}$ and their longitudinal momentum $p_3$
given by $\Delta p_3=-gE\Delta t$; it is owing to the acceleration
by the electric field with the interval $\Delta t=\sqrt{gB}/gE$.
Among them, the mode with the vanishing transverse and
longitudinal momenta ( $p_T=0$ and $p_3+gA_b=0$ ), has the largest amplitude
and growth rate.
Furthermore, as we will show below, 
the amplitude of the mode can grow up to of the order of $1/g$
until the nonlinear interactions among the unstable modes are effective.
Hence, we may think that the mode form a Bose condensate of gluons with zero momentum.

We have classically discussed the gluon production in the previous sections. 
It means that
the formulae may be regarded as the ones concerning the gluon condensate.
We now show using the formulae that the number density $n_g$ of the gluons produced 
is of the order of $Q_s^3/g^2$, i.e. $n_g\sim Q_s^3/g^2$ for small $g\ll 1$. 
The most relevant mode for the production is the one with zero momentum $p_3=0$,
as shown in the Fig 2. As long as $gA_b \ll \sqrt{gB}=Q_s$,
we can approximately obtain a solution 
$\tilde{\phi}(0)=\exp(Q_s\,t)$ from the second equation with $p_3=0$ in eq(\ref{eq}).
The solution holds near $t=0$
because of the initial condition $gA_b(t=0)=0$.
The solution approximately holds until $gA_b$ is comparable with $Q_s$.
In order to see how $gA_b$ increases with time,
we rewrite the Maxwell equation in eq(\ref{eq}) such that

\begin{equation}
\partial_0^2 gA_b=-\frac{g^2\sqrt{gB}}{4\pi}\int_{-\sqrt{gB}}^{+\sqrt{gB}} 
dp_3 \frac{(p_3+gA_b)\tilde{\phi}(p_3)^2}{T(p_3l_B)}\sim -\frac{g^2\sqrt{gB}}{4\pi}
\frac{2\sqrt{gB} \,gA_b\,\tilde{\phi}(0)^2}{T(0)}=-g^2\frac{Q_s^2\,\tilde{\phi}(0)^2gA_b}{2\pi T(0)} ,
\end{equation}
with $T(0)=(\sqrt{\pi/2})/4$,
where we neglected the contribution of the quarks.
As long as $g^2\tilde{\phi}(0)^2$ is much small, the electric field $E=g\partial_0A$
slowly decreases. But, 
once $g^2\tilde{\phi}(0)^2$ reaches of the order of $1$, the electric field 
$E=g\partial_0A_b$ rapidly decreases and vanishes. 
Thus, we may approximately estimate the life time $t_c$ of the electric field such that
$g^2\tilde{\phi}(0)^2=g^2\exp(Q_st_c)=1$, i.e. $t_c=-Q_s^{-1}\log(g^2)$.
In this way we find that $\tilde{\phi}(0,t=t_c)^2\sim g^{-2}$ and $gA_b\sim gE_0t_c\sim Q_s$ 
when the electric field vanishes.
The unstable mode $\phi\propto \tilde{\phi}(0)$ grows up to of the order of $g^{-1}$.
Using the result we can show that the number density of the condensed gluons 
is of the order of $g^{-2}$.
Here we remember that the current carried by the four kinds of massless quarks is given by $4gn_q$,
where $n_q$ denotes the number density of a kind of the quarks.
Thus, we may define the number density $n_g$ of gluons such as $n_g=J_g/g$
in the classical approximation. Then, it follows that

\begin{equation}
n_g=\frac{gB\tilde{\phi}(0,t=t_c)^2gA_b(t)}{2\pi T(0)}\sim \frac{Q_s^3}{g^2}.
\end{equation}

Therefore, we find that the gluon condensate 
arises owing to the anomalous gluon production and that
the number density of the gluons increases up to of the order of
$Q_s^3g^{-2}$. The result has been expected\cite{condense} when
the number of gluons is conserved, in other words, nonlinear interactions
violating the gluon number conservation is not effective.
The expectation comes from the fact that
the number density of the gluons forming the background gauge fields $A_b$ is of the 
order of $Q_s^3g^{-2}$, 
while the number density of thermalized gluons produced by the decay of $A_b$ 
is of the order of $Q_s^3$.
( The energy density $\epsilon_g$ of the thermalized gluons is given 
such as $\epsilon_g \propto n_g^{4/3}$ in terms of $n_g$ or 
$\epsilon_g \propto Q_sn_g$ since the typical energy of the gluon 
is $Q_s$. Thus, we find that $n_g\propto Q_s^3$. )
In order to derive our result, we simply use the fact that
the initial magnitude of the unstable modes is of the order unity. 
Thus, our result does not depend on the detail of the initial 
conditions for the unstable mode $\tilde{\phi}(0,t=0)$ used in our paper.

\vspace{0.1cm} 
In our calculations we have not included interactions among the condensate, namely,
we have neglected the four point interactions of $\Phi$.
( The field $\Phi$ involves modes 
in higher Landau levels as well as the unstable modes in the lowest Landau level. )
Once the interactions are effective, the condensate would melt.
Contrary to the expectation from the similarity to the model of the complex scalar field,
the nonlinear interactions in the gauge theory do not stabilize
the condensate.
The interactions cause the momentum transfer 
from the mode $\phi$ to the other stable modes in higher Landau levels
or they produce new type of unstable modes.
Indeed,
as we have shown in the previous paper\cite{instability2}, 
seconday Nielsen-Olesen unstable modes are
induced after the primary unstable modes $\phi$ grow sufficiently large
for the nonlinear interactions to be effective. That is, 
the localized electric currents of each $\phi_l$ in eq(\ref{sum})
become large and induce a magnetic field surrounding
the current of $\phi_l$. We called it azimuthal magnetic field in the paper.
Under the azimuthal magnetic field, the secondary Nielsen-Olesen unstable modes are induced.
The modes carry\cite{ven} larger momentum $p_3$ than $Q_s$, while the unstable modes primarily induced
carry smaller momentum than $Q_s$. 
The primary unstable modes form the Bose condensate with zero momentum,
which induces the secondary unstable modes with large momentum.
In this way, 
the momentum transfer occurs from the condensate to
the secondary unstable modes.
As a result the condensate would melt. 
The excitations of the secondary unstable modes are caused by
the nonlinear interactions. 
The result has also been expected
in the reference\cite{condense}.
We would like to point out that the cascade from small momentum to
large momentum shown above
can be seen in a model of scalar fields \cite{berges}.

We can estimate when the nonlinear interactions are effective. 
We note that
the nonlinear interactions are given schematically by $g^2\Phi^4$ or $g(A_b+\delta A)\Phi^2$ 
in eq(\ref{L}); the background gauge fields $A_b$ are assumed to be of the order of $1/g$.
The interactions $g^2\Phi^4$ or $g\delta A\Phi^2$ are smaller than the kinetic terms of $\Phi$ and $A_b$
if the amplitude of the unstable mode $\phi$ or $\Phi$ is much less than of the order of $1/g$. 
But when $\Phi$ reaches of the order of $1/g$, 
all of the interaction terms become 
of the same order of the magnitude as the kinetic terms.
( Note that the fluctuation  $\delta A$ is proportional to the term $g\Phi^2$ in the equation of motion
of $\delta A$ and becomes the same order of magnitude as $\Phi$ when $\Phi\sim g^{-1}$. )
Hence, the nonlinear interactions become effective when $\Phi\sim g^{-1}$ or
the number density of the condensate reaches of the order of $g^{-2}$.

Consequently, owing to the anomalous gluon production, 
the Bose condensate appears in the process of the decay of 
the background gauge fields $E$ and $B$. 
However, the condensate melts 
after the number density of the gluons in the condensate grows up to
of the order of $g^{-2}$. Eventually, thermalized QGP would be realized.

\section{summary and discussion}
\label{6}

Motivated by the recent study of
the anomalous production of the Nielsen-Olesen unstable modes 
by the Schwinger mechanism,
we have discussed the decay of the color electric field
in the classical approximation by taking into account the back reaction of the electric
field to the gluon and quark production. We have found that
the electric field rapidly decays
owing to the anomalous production of the gluons. 
It has turned out that the contribution of the quarks to the decay is negligible.
We have also found that the amount of the produced gluons is about a hundred times larger
than that of the quarks.

A model of color glass condensate predicts that the color electric and magnetic fields
are produced immediately after high energy heavy ion collisions. 
Fluid dynamical simulations of thermalized QGP
suggest that
the fields should decay into the plasma within the time $1\,$fm/c.
Our analysis indicates that such a very fast decay is caused 
by the anomalous gluon production. Actually, our analysis shows that 
the decay is completed within the time of the order of $Q_s^{-1}$.
 
We have also shown that
the gluons of the Nielsen-Olesen unstable mode form the Bose condensate
with zero momentum. 
The number of the gluons in the condensate rapidly increases and 
is saturated when it becomes of the order of $1/g^2$. After the saturation,  
the nonlinear interactions are effective so that 
the rapid momentum transfer from the condensate to modes
with large momentum ( $>\sqrt{gB}$ ) arises. Hence the equipartition of the momentum 
and the thermalization of QGP
would be achieved.

We have discussed the decay of the color electric field in the glasma. 
The color magnetic field in the glasma also decays in the following.
In general the longitudinal color magnetic fields form flux tubes, 
which expands with time. Owing to the expansion, electric field $\delta E_T$
perpendicular to the magnetic field is induced 
according to the Faraday's law of induction. On the other hand,
owing to the expansion of the longitudinal electric flux tubes,
magnetic field $\delta B_T$ is induced, which is parallel or anti-parallel 
to the electric field $\delta E_T$.
Thus, under the field $\delta B_T$, Nielsen-Olesen unstable modes $\delta \phi_T$ are excited and 
make the electric field $\delta E_T$
decay rapidly. Eventually, the expansion of 
the magnetic flux tube induces the electric field $\delta E_T$,
which decays owing to the acceleration of the unstable modes $\delta \phi_T$.
This implies the decay of the magnetic flux tube.
This is the decay mechanism of the magnetic field.

We have discussed the gluon production in the non-expanding glasma.
When we treat it in the expanding glasma, the similar analysis is possible.
But, we should use initial conditions for the unstable modes 
shown in the recent paper\cite{ven2}.
We can show that the "free fluctuations"  
in the reference are identical to the vacuum fluctuations
in our discussion when the "free fluctuations" are formulated in the Cartesian coordinates.
On the other hand, relevant fluctuations are "fluctuations in the Glasma".
Thus, the initial condition in our paper apparently seems to be not appropriate.
But, the fluctuations in Glasma are typically represented by  
field configurations in the lowest Landau level. 
We have used such a typical field configuration 
as the initial condition.  
In that sense, our choice of the initial condition
is not necessarily inappropriate, 
although the longitudinal momentum distribution in the initial condition
is different from the ones
of the "fluctuations in the Glasma".   
In both cases, the magnitudes of the initial unstable modes are of order unity. 
Our results do not depend on the detail of the initial conditions.
Thus, our results might hold in general.
We wish to discuss the decay of glasma by using
initial conditions dictated in the reference\cite{ven2}
in the near future.

\vspace*{1cm}
We would like to
express thanks to Dr. K. Fukushima for making me pay attention to the recent paper
by Berges and Sexty.


\end{document}